\documentclass[preprint,nofootinbib,longbibliography]{revtex4-1}
\usepackage[utf8]{inputenc}
\usepackage{amsmath}
\usepackage{amsfonts}
\usepackage{xcolor}
\usepackage{slashed}
\usepackage{mathtools}

\usepackage[colorlinks=true]{hyperref} 
\hypersetup{
    bookmarks=true,         
    unicode=false,          
    pdftoolbar=true,        
    pdfmenubar=true,        
    pdffitwindow=false,     
    pdfstartview={FitH},    
    pdfsubject={},   
    pdfcreator={},   
    pdfproducer={}, 
    pdfkeywords={} {} {}, 
    pdfnewwindow=true,      
    colorlinks=true,       
    linkcolor=magenta, 
    citecolor=blue,        
    filecolor=magenta,      
    urlcolor=blue           
}

\newcommand{\rhob}{\overline \rho}
\newcommand{\psib}{\overline \Psi}

\begin{document}
\count\footins = 1000
\title{A note on GMP algebra, dipole symmetry, and Hohenberg-Merming-Wagner theorem in the lowest Landau level. }
\begin{abstract}
    After projection to the lowest Landau level  translational invariance and particle conservation combine into dipole symmetry. We show that the new symmetry forbids spontaneous $U(1)$ symmetry breaking at zero temperature. In the case of the spatially inhomogeneous magnetic field, where the translational invariance is absent, we show that the dipole symmetry disappears and the constraint on the symmetry breaking is lifted. We pay special attention to the fate of the Girvin-Macdonald-Platzman algebra in the inhomogeneous magnetic field and show that a natural generalization of it is still present even though the dipole symmetry is not. 
\end{abstract}
\author{Lev Spodyneiko}
\email{lionspo@mit.edu}
\affiliation{Massachusetts Institute of Technology, Cambridge, MA 02139, United States}
\maketitle
\section{Introduction}

In a strong magnetic field, the domination of the interactions over kinetic energy leads to rich strongly correlated physics. The importance of many-body effects complicates the theoretical understanding of the phase diagram of these systems. It was proposed \cite{KS} recently that some insight can be achieved by considering toy models with dipole symmetry where both $U(1)$ charge and its dipole moment are conserved. The dipole moment conservation leads to restricted mobility of charged excitations, mimicking the quenching of kinetic energy in the magnetic field. 

It was shown \cite{KS} that dipole symmetry leads to the absence of spontaneous $U(1)$-symmetry breaking at zero temperature in two-dimensional systems. This is an extension of the Hohenberg-Mermin-Wagner (HMW) theorem \cite{Hohenberg,MW} to this setting. An important caveat is that the dipole symmetry algebra used in the proof did not involve the effects of the non-trivial topology of the wave functions. In the language of the band theory, this would mean that the Berry curvature and magnetic field are zero.  This is certainly not a good approximation even for the simplest case of the spatially homogeneous magnetic field.

The primary purpose of the paper is to show that in the presence of a homogeneous magnetic field the results of \cite{KS} still hold. More precisely, we take the limit where the cyclotron energy is much larger than the typical interaction scale and  restrict the system to the lowest Landau level (LLL). In this limit, a dipole symmetry emerges: not only the projected charge is conserved but also its dipole moment. The additional conservation law modifies the f-sum rule which in turn forbids the spontaneous $U(1)$-symmetry breaking in the LLL at zero temperature. The same conclusion was made in \cite{Pitaevskii1991} by L. Pitaevskii
and S. Stringari. In the present paper, we interpret their result using dipole symmetry and also discuss how it generalizes to the inhomogeneous magnetic field.

In recent years, the question of whether the lowest Landau level physics and fractional quantum Hall effect (FQHE) in particular can be reproduced in the absence of a magnetic field attracted  a lot of attention. For example, one can expect to see FQHE state in a flat partially filled Chern band. Experimental signatures of such fractional Chern insulating states were reported in \cite{FCIingraph}. However, one still lacks a coherent theoretic understanding of how these states can emerge. The LLL has a lot of fascinating features and figuring out which of them are essential for the existence of FQHE phases is an open problem. 


In this paper, we will take an opportunity to shed some light on this problem by focusing on the Girvin-Macdonald-Platzman (GMP) algebra \cite{GMP} which was discussed in this context in \cite{GMPinBands, GMPinBands2} (see also \cite{tracecond,vortex}). As the main example, we will consider Landau levels in a spatially varying field, which is a toy model for a flat band that satisfies the trace condition \cite{tracecond}, has non-trivial Chern class, and an inhomogeneous Berry curvature across the Brillouin zone. A well-known fact in K\"ahler quantization literature (see for example Proposition 4.1 in \cite{kahlerquant}) is that the projected density operators form a closed algebra. Though the structure constants of the resulting algebra are different from the homogeneous case, it is still a natural generalization of the GMP algebra. Despite the presence of the algebra, the physically important property of translational invariance  (and dipole symmetry) is absent leading to changes in the behavior such as the inapplicability of the HMW theorem.

The plan of the paper is as follows. In Section \ref{sec: LLL} we  review the lowest Landau level projection using physical space instead of reciprocal space which is more common in the literature. In section \ref{sec: HMW} we will prove that $U(1)$ symmetry cannot be broken in the LLL assuming that the magnetic field is spatially homogeneous. In section \ref{sec: no homo}, we will consider a simple model of spatially varying magnetic field and compute its GMP algebra. We will explain how it invalidates the proof of the HMW theorem of Section \ref{sec: HMW}. We will discuss the limitations of the argument and possible future directions in the last section. 

We thank   Anton Kapustin, Patrick J. Ledwith, Sergej Moroz, Dam Thanh Son, and Senthil Todadri  for discussions. The work was supported by the Simons Collaboration on Ultra-Quantum Matter, which is a grant  (651446) from the Simons Foundation.

\section{LLL projection} \label{sec: LLL}
In this section, we will introduce the lowest Landau level projection. The procedure is rather standard by now but we will use a slightly different perspective on it. We want to use the procedure in a spatially varying magnetic field. In this case, the continuous translational symmetry is absent and thus the use of position space notation instead of momentum space notation is more appropriate. This is essentially a generalization of appendix A in \cite{Readboson} to the inhomogeneous magnetic field. The main point is that even though the GMP algebra of density operators is still closed in an inhomogeneous field, the dipole symmetry is lost. Since it is the dipole symmetry that forbids the $U(1)$-symmetry breaking, the latter can be spontaneously broken in an inhomogeneous field. More generally, one can consider electors  propagating on a curved manifold in the presence of an inhomogeneous magnetic field \cite{curvedLLL}. After the transition to isothermal coordinates, the problem essentially reduces to the flat space and we will restrict ourselves to the latter for simplicity of presentation. All results of this section can be generalized to a curved background with suitable adjustments. 

Consider a Hamiltonian of interacting fully spin-polarised fermions confined to a plane in an inhomogeneous magnetic field
\begin{align*}\label{eq: hamiltonian}
    H= \int d^2{\bf x} \Big[\,\Psi^\dagger({\bf x})&\frac{(\hat {\bf p}-{\bf A} )^2}{2m}\Psi({\bf x})- \frac{g}{4m} B({\bf x})\rho({\bf x})\Big]+ \frac 1 2 \int d^2{\bf x}d^2{\bf y}\,V({\bf x}-{\bf y})\rho({\bf x})\rho({\bf y}),
\end{align*}
where  $\rho({\bf x})=\Psi^\dagger({\bf x}) \Psi({\bf x})$ is a density operator, $\bf A$ is a vector potential of the magnetic field $B(\bf x)$ pointed perpendicular to the plane. We have included the Zeeman term and ignored the other spin polarization since we will focus only on the LLL where all particles are spin-polarized. The creation-annihilation operators satisfy canonical anti-commutation relations 
\begin{align}
    \{\Psi({\bf x}), \Psi^\dagger({\bf y})\} = \delta({\bf x - y}),
\end{align}
with other anti-commutators being zero.

By introducing the operator 
\begin{align}\label{eq: cov derivative def}
    D = (\hat p_1-A_1)+i(\hat p_2-A_2),
\end{align}
we can rewrite the Hamiltonian as
\begin{align*}\label{eq: hamiltonian in holomorphic form}
    H= \int d^2{\bf x} \Big[\,\Psi^\dagger({\bf x})&\frac{D^\dagger D}{2m}\Psi({\bf x})- \frac{g-2}{4m} B({\bf x})\rho({\bf x})\Big]+ \frac 1 2 \int d^2{\bf x}d^2{\bf y}\,V({\bf x}-{\bf y})\rho({\bf x})\rho({\bf y}).
\end{align*}
Let us first ignore the interactions and focus on the single-particle part of the Hamiltonian. After setting the $g$-factor\footnote{ When $g=2$ the Paul Hamiltonian is equal to a square of 2d Dirac operator. In a magnetic field, the Dirac operator conserves chirality and has chiral zero modes which are protected by the Atiyah-Singer index theorem.  Alternatively, one can motivate $g=2$ by considering a slowly varying magnetic field. In this case, the first term of the Hamiltonian will have zero-point energy (equal to half of the cyclotron frequency) which will depend on the position. The Zeeman term with $g=2$ exactly cancels this contribution.} to $g=2$, the non-interacting particle's ground state is degenerate and is spanned by functions satisfying $D\psi({\bf x})=0$. It is a natural generalization of the lowest Landau level to the case of an inhomogeneous magnetic field. We will assume that the flux of the magnetic field averaged over the whole space is positive. In this case, the LLL is not empty and the number of states in it scales as the total flux divided by $2\pi$. For a generic magnetic field configuration and positive average flux, the first excited non-interacting level will be separated by a gap proportional to inverse mass $\frac 1 m$. By taking the limit $m\rightarrow 0$, we can neglect higher Landau levels and project the dynamics to the lowest Landau level. On the physical grounds, it should be a good approximation in the limit when typical interaction energy is less than the cyclotron frequency.   We will denote the full single-particle Hilbert space as $\mathcal H$ and the ground state subspace as $\mathcal H_{\rm LLL}$.

Let $\psi_n({\bf x})$ be an orthonormal basis in the LLL enumerated by $n\in \mathbb N$. Consider the following function 
\begin{align}
    K({\bf x},{\bf y}) = \sum_{n\in \mathbb N} \psi_n({\bf x}) \psi_n^\dagger(\bf{y}).
\end{align}
It is a kernel of the projection $\hat K$ to the LLL and satisfies a number of useful properties
\begin{align}
  K({\bf x},{\bf y}) &= \int d^2{\bf z}  \, K({\bf x},{\bf z})   K({\bf z},{\bf y}),  && &&(\text{idempotnence})\\
 K({\bf x},{\bf y})^* &= K({\bf y},{\bf x}), && &&(\text{hermicity})\\
 D_{\bf x} K({\bf x},{\bf y})&= 0, && &&(\text{holomorphicity in $\bf x$})\\
 D_{\bf y}^\dagger K({\bf x},{\bf y})&= 0, && &&(\text{antiholomorphicity in $\bf y$})\\
 \psi({\bf x}) &= \int d^2 {\bf y} \,  K({\bf x},{\bf y}) \psi({\bf y}), \quad \text{iff $\psi\in\mathcal H_{\rm LLL}$}, && &&(\text{reproducing property})
\end{align}
where $D_{\bf x}$ is the operator (\ref{eq: cov derivative def}) acting only on $\bf x$. Note, that $D$ satisfies the integrability condition $[D,D]=0$  and therefore can be used to define complex coordinate patches on the $U(1)$ bundle in which $\psi({\bf x})$ lives. In this coordinate system, $D$ becomes an anti-holomorphic derivative and the function  $K({\bf x},{\bf y})$ becomes holomorphic in $\bf x$ and antiholomorpohic in $\bf y$. In particular, it is uniquely determined by its diagonal $K({\bf x},{\bf x})$. Note, we are not working in a specific gauge (e.g. symmetric gauge), and in our approach, the complex charts are defined by the connection $D$. In a symmetric gauge, our notion of holomorphicity reduces to the usual one. 

An important fact that we will use extensively in the following is that any single-particle operator on the projected Hilbert space can be represented as a projection of the operator of multiplication by a function. Denote, the projected operators as $\hat T_f=\hat K \hat f \hat K$, where $\hat f$ is an operator of multiplication by a function $\hat f \psi({\bf x}) = f({\bf x}) \psi ({\bf x})$ in the unprojected single-particle Hilbert space. We will occasionally call operators of this form Toeplitz operators. Suppose that there is an operator $\hat A$ which is orthogonal (in the Hilbert-Schmidt sense) to $\hat T_f$ for all $f$, then
\begin{align}\label{eq: surjectivity}
  0 &= {\rm tr}(\hat A \hat T_f) = \int d^2{\bf x}\, A({\bf x},{\bf x}) f({\bf x}),
\end{align}
where we defined $A({\bf x},{\bf y})= \sum_{n,m} \psi_n({\bf x})\langle n | \hat A|m\rangle \psi_m^\dagger({\bf y})$. This function is holomorphic in $\bf x$ and antiholomorphic in $\bf y$. The equation above means\footnote{The left-hand and right-hand sides of (\ref{eq: surjectivity}) must be well-defined for this argument to work. This requires the operator $\hat A$ to be compact. One can refine the argument by replacing the Hilbert-Schmidt topology with the strong operator topology, i.e. by considering matrix elements instead of the trace. The precise correspondence in strong operator topology is not between all operators on  $\mathcal H_{\rm LLL}$ and space of all functions on $\mathbb R^2$  but between bounded operators and essentially bounded functions. See \cite{Densityoftoeplitz} for more details.} that $A({\bf x},{\bf x}) =0$. By analytic continuation $A({\bf x},{\bf y})=0$ and so is  $\hat A$. 

On a plane, the equality of operators $\hat T_f = \hat T_g$ leads to the equality of functions $f=g$ (see e.g. Proposition 2.95 in \cite{HarmonicAnal}). If we compactify the system on a torus, this is non-longer true and the map $f\rightarrow \hat T_f$ will have a non-trivial kernel. The representation of operators by functions should be understood modulo the kernel in this case.  

Now, we can turn to the many-body Hamiltonian. We can introduce the projected field as
\begin{align}
    \psib({\bf x}) &= \int d^2{\bf y}  K({\bf x},{\bf y})\Psi({\bf y}).
\end{align}
In terms of the mode expansion $\Psi({\bf x}) = \sum_{\psi_n\in \mathcal H} \psi_n({\bf x})c_n$, the projected operator $\psib({\bf x}) = \sum_{\psi_n\in \mathcal H_{\rm LLL}} \psi_n({\bf x})c_n$  contains only modes in the LLL. It satisfies
\begin{align}
    \{\psib ({\bf x}),\psib^\dagger({\bf y})\} &= K({\bf x}, {\bf y}),\\
  \{\psib ({\bf x}),\psib({\bf y})\} &= 0,\\
  \{\psib^\dagger ({\bf x}),\psib^\dagger({\bf y})\} &= 0,\\
    D\psib({\bf x}) = 0. \label{eq: hol of psib}
\end{align}

We will denote the projection of the many-body Hilbert space to the subspace where particles occupy only the lowest Ladnau level by $P_{\rm LLL}$. After these preliminaries the projection of the many-body Hamiltonian is straightforward. One has to normal order it (with respect to $\Psi^\dagger, \Psi$), then replace $\Psi \rightarrow \psib$ and undo the normal ordering. Normal ordering excludes processes where particles jump to higher Landau levels and then return back to LLL. For example, 
\begin{align*}
    P_{\rm LLL} \rho({\bf x)}\rho({\bf y})  P_{\rm LLL}&= -P_{\rm LLL} \Psi^\dagger ({\bf x)}\Psi^\dagger ({\bf y)} \Psi ({\bf x)}\Psi ({\bf y)}  P_{\rm LLL} +\delta({\bf x - y}) P_{\rm LLL} \Psi^\dagger({\bf x})\Psi({\bf y})P_{\rm LLL} \\ &=  - \psib^\dagger ({\bf x)}\psib^\dagger ({\bf y)} \psib ({\bf x)}\psib ({\bf y)}   +\delta({\bf x - y}) \psib^\dagger({\bf x})\psib({\bf y}) \\&=  \rhob({\bf x})\rhob({\bf y}) +(\delta({\bf x - y})-K({\bf x},{\bf y}))\psib^\dagger({\bf x})\psib({\bf y}),
\end{align*}
where we have defined the projected density operator $\rhob({\bf x}) = \psib^\dagger ({\bf x}) \psib ({\bf x})$.
Due to the holomorphicity (\ref{eq: hol of psib}) of $\psib$ the kinetic term projects to zero. 

The resulting projected Hamiltonian is 
\begin{align*}
    \overline  H =  \frac 1 2 \int d^2{\bf x}d^2{\bf y}V({\bf x}-{\bf y})  \rhob({\bf x})  \rhob({\bf y}) &+\frac 1 2  \int d^2{\bf x} V({\bf 0}) \rhob({\bf x}) \\&- \frac 1 2  \int d^2{\bf x}d^2{\bf y}V({\bf x}-{\bf y}) K({\bf x},{\bf y})\psib^\dagger({\bf x})\psib({\bf y}).
\end{align*}
According to the discussion above, the last term can be rewritten as a projection of a single-particle potential
\begin{align}
     - \frac 1 2  \int d^2{\bf x}d^2{\bf y}V({\bf x}-{\bf y}) K({\bf x},{\bf y})\psib^\dagger({\bf x})\psib({\bf y}) =  \int d^2{\bf x} \,\phi({\bf x}) \rhob({\bf x}),
\end{align}
where the potential $\phi({\bf x})$ depends on interactions $V({\bf x}-{\bf y})$ and the kernel $K({\bf x},{\bf y})$. If the magnetic field is homogeneous the resulting function $\phi({\bf x})$ will be constant which leads to a simple shift of the chemical potential. However, for inhomogeneous field $\phi({\bf x})$ will have a non-trivial spatial dependence. This may appear counter-intuitive. Indeed consider a Landau level with just one particle. The last term in the Hamiltonian suggests that the interaction will generate a spatially non-trivial Hamiltonian even though the particle has nothing to interact with. The resolution of this problem is that the first term is also non-zero even for a single particle and it is exactly compensated by the last term.

Let us define a projection of a density operator weighted by a function $f({\bf x})$
\begin{equation}
    \rhob(f) = P_{\rm LLL}\int d^2 {\bf x} \,f({\bf x})\rho({\bf x}) P_{\rm LLL} = \int d^2 {\bf x} \,f({\bf x})\psib^\dagger({\bf x})\psib({\bf x}).
\end{equation}
The commutator of two such operators is
\begin{align} \label{eq: star product def}
    [\rhob(f),\rhob(g)] = \int d^2 {\bf x}d^2 {\bf y} \,\psib^\dagger({\bf x}) \big[f({\bf x}) K({\bf x},{\bf y}) g({\bf y})-f({\bf y}) K({\bf y},{\bf x}) g({\bf x})\big]\psib({\bf y}).
\end{align}
The r.h.s. is a single particle operator. As explained above, all single-particle operators are of the form $\rhob(h)$ for some function $h$. Therefore we can express the r.h.s. as $h = f\star g -g\star f$, where the star product is implicitly defined by the equation $\hat T_f \hat T_g = i\hat T_{f\star g}$. We find, 
\begin{align} \label{eq: star product def 2}
    [\rhob(f),\rhob(g)] =i \rhob(f\star g-g\star f).
\end{align}
This is a natural generalization of GMP algebra to the case of the inhomogeneous magnetic field. One can see that the projected Hamiltonian can be expressed in terms of the density operators $\rhob$ which form a closed algebra. Unfortunately, explicit formulas for the star product require knowledge of the kernel $K({\bf x},{\bf y}) $ which is only computable for special cases. We will discuss an example with an inhomogeneous magnetic field and compute the resulting GMP algebra in Section \ref{sec: no homo}.

Suppose the Hamiltonian is invariant under a unitary transformation $U$
\begin{equation}
U H U^{-1} =H.
\end{equation}
Commutativity with the Hamiltonian means that even after the projection the operator  $P_{\rm LLL} U P_{\rm LLL}$ stays unitary and commutes with the projected Hamiltonian. We will assume that a single-particle operator conjugated by $U$ stays a single-particle operator. In particular
\begin{align}\label{eq: sym action on rho}
    U \rhob(f) U^{-1} = \rhob(f^U)
\end{align}
for some function $f^U$. Then the star product is covariant under the symmetry
\begin{align}
    (f*g)^U = f^U\star g^U.
\end{align}
Note, the action of symmetry $U$ on functions can be determined from the action of $U$ on unprojected operators $\rho(f)$. For example, magnetic translation \cite{Zak} operators  $\mathcal T_{\bf a}$ act as $f^{\mathcal T_{\bf a}}({\bf x}) = f({\bf x+a})$.

\subsection*{Spatially homogeneous magnetic field}
 Let us now focus on the homogeneous magnetic field $B({\bf x})=B_0$. In this case, the algebra of the projected densities $\rhob(f)$ is covariant under the magnetic translation operators $\mathcal T_{\bf a}$. It is natural to expand the projected densities in terms of eigenvectors of the operator $\mathcal T_{\bf a}$ acting on $f$. We define
 \begin{align}
     \rhob_{\bf k} = \rhob(e^{-i {\bf k \cdot x}}),
 \end{align}
 where minus on the right-hand side is to match the standard definition of the Fourier transform.

Magnetic translations do not commute with each other and one cannot diagonalize translations in different directions simultaneously. However, the density operators transform under the adjoint representation (\ref{eq: sym action on rho}) of magnetic translation algebra in which the translations do commute. Thus one can expand $\rhob$ in terms of usual plane waves.

 The GMP algebra in momentum space\footnote{Note that the right-hand side contains exponentially large structure constant. In position space, it will lead to an algebra of $\rhob(f)$ that is well-defined only for analytic functions $f$ on the whole plane. This does not lead to any problems since one can replace arbitrary function $f$ with its convolution with sharp Gaussian function $\widetilde f ({\bf x})= \int{d^2{\bf y}}\frac 1 {{2\pi \varepsilon}}\exp(-({\bf x}-{\bf y})^2/2\varepsilon) f({\bf y})$. The resulting operator $\hat T_{\widetilde f}$ will converge to $\hat T_{f}$ as $\varepsilon\rightarrow0$. The GMP algebra can be computed directly in the position space. Intermediate steps in the computation require finding an inverse to a heat kernel, which can be found in \cite{inverseheat}.

 }
 \begin{align}\label{eq: GMP}
    [\rhob_{\bf k},\rhob_{\bf q}] = 2i\, \exp \Big({\dfrac {{\bf k}\cdot {\bf q}} 2l^2_B} \Big)\sin  \Big(\frac { {\bf k}\wedge {\bf q}} 2 l^2_B\Big)\, \rhob_{{\bf k}+{\bf q}},
 \end{align}
where $l_B=\sqrt{\frac 1 {B_0}}$ is the magnetic length, $(\wedge {\bf  k})_i = \epsilon_{ij}k_j$ with $\epsilon_{ij}$ being the Levi-Civita symbol, ${\bf k}\wedge {\bf p} = {\bf k} \cdot (\wedge {\bf p})$, and summation over repeating indices is assumed. By expanding the algebra at small $\bf k$, we see that infinitesimal translations are generated by the operator $P_j = il_B^{-2}\epsilon_{jm} \dfrac{\partial{\rhob_{\bf k}}}{\partial k_m} \Big|_{{\bf k}=0}= l_B^{-2}\epsilon_{jm}\rhob(x_m)$
\begin{equation}
    [P_i, \rhob_{\bf k}] =  k_i \rhob_{\bf k},
\end{equation}
or in position space
\begin{align}
    [P_i, \rhob({\bf x})] =  \frac{\partial \rhob({\bf x})}{\partial x_i}.
\end{align}

Let us define the total charge $Q=\rhob(1)=\int d^2{\bf x}\, \rhob({\bf x})$ and the dipole charge $D_i = \rhob(x_i)= \int d^2{\bf x}\, x_i \rhob({\bf x}) $. Note, that dipole and momentum operators are related to each other $D_i=-l_B^2\epsilon_{ij} P_i$. Importantly, the densities of $D_i$ and $Q$ are related as $d_i({\bf x}) = x_i q({\bf x}) $, i.e. $D_i$ can be thought as a dipole moment of the charge $Q$. Therefore, the system projected to the lowest Landau level conserves not only the charge but also the associated dipole moment.

\subsection*{Periodic magnetic field}

Lastly, we will consider a periodic magnetic field $B({\bf x}+{\bf R})=B({\bf x})$ for all $\bf R$ which belong to a lattice $\Lambda$. In this case, the magnetic translations $\mathcal T_{\bf R}$ by the lattice vectors ${\bf R} \in \Lambda$ act as a symmetry. 

As in the homogeneous case, the projected densities $\rhob(\bf x)$ transform in the adjoint representation of the magnetic translation group which is isomorphic to the adjoint representation of the translation group in the absence of a magnetic field. Thus, it is natural to decompose an arbitrary function $f$ as
\begin{align}
    f({\bf r}) = \sum_{n}\int_{{\bf K} \in {\rm BZ}} \frac{d^2{\bf K}}{(2\pi)^2} f_{{\bf K},n} e^{-i {\bf K} \cdot {\bf r}} g_n({\bf r}),
\end{align}
where $g_n({\bf r})$ is a basis of periodic functions $g_n({\bf r}+{\bf R})=g_n({\bf r})$ for all ${\bf R} \in \Lambda$, and $\bf K$ is integrated over first (non-magnetic) Brillouin zone. There is no {\it a priori} preferred choice of the basis $g_n({\bf r})$. 

Define
\begin{equation}
    \rhob_{{\bf K},n} =  \rhob(e^{-i {\bf K \cdot {\bf r}}} g_n({\bf r})).
\end{equation}
The GMP algebra will have the form
 \begin{align}
[ \rhob_{{\bf K},n}, \rhob_{{\bf Q},m}] = \sum_l C_{nm}^l({\bf K},{\bf Q})\, \rhob_{{\bf K}+{\bf Q},l},
 \end{align}
 where ${\bf K}+{\bf Q}$ is understood modulo reciprocal lattice vectors ${\bf G} \in \Lambda^{\vee}$. The structure constants $C_{nm}^l({\bf K},{\bf Q})$ are complicated functionals of the magnetic field $B({\bf x})$ and its computation is abstracted by the absence of explicit expression for the kernel $K({\bf x},{\bf y})$. 

 As was mentioned above, one is free to choose any basis for $g_n({\bf x})$. A simple choice is $g_{\bf G}({\bf x}) = \exp(-i {\bf G}\cdot {\bf x})$, where index $n$ is replaced by a vector lying in the reciprocal lattice ${\bf G} \in \Lambda^{\vee}$. We can combine $\bf K$ and $\bf G$ into a single vector ${\bf k}={\bf K}+{\bf G}$ which is not restricted to the first Brillouin zone
 \begin{equation}
    \rhob_{{\bf k}}= \rhob_{{\bf K},{\bf G}} =  \rhob(e^{-i {\bf k \cdot {\bf r}}} ).
\end{equation}
In this basis, the GMP algebra has the form
\begin{align}
    [\rhob_{{\bf k}},\rhob_{{\bf q}}] =\sum_{{\bf G} \in \Lambda^{\vee}} C^{{\bf G}}({\bf k},{\bf q}) \rhob_{{\bf k} + {\bf q}+{\bf G}}.
\end{align}
 One should note that the GMP algebra does not appear to be sensitive to the central extension of the translational group to the magnetic translation group. The above discussion work equally well for integer, rational, or even irrational total flux through the unit cell. In contrast, the structure of the wave functions and magnetic Brillouin zone strongly depends on the number theoretic properties of the flux \cite{DubNov}. It is an interesting question whether something special happens to the structure constants $C^{{\bf G}}({\bf k},{\bf q})$ at rational or integer flux through unit cell. 
\section{Hohenberg-Mermin-Wagner theorem}\label{sec: HMW}
In this section, we prove the absence of spontaneous $U(1)$ symmetry breaking in a system of particles in the LLL with density-density interaction and in the presence of a spatially homogeneous magnetic field. 

The Hamiltonian is
\begin{align}
    \overline H= \frac 1 2 \int d^2{\bf x}d^2{\bf y}\,V({\bf x}-{\bf y})\rhob({\bf x})\rhob({\bf y}) = \frac 1 2 \int \frac{d^2{\bf q}}{(2\pi)^2} V_{\bf q} \rhob_{-\bf q}\rhob_{\bf q},
\end{align}
where we have ignored the chemical potential shift, $V_{\bf q}$ is Fourier transform of $V({\bf x}-{\bf y})$, and projected density operators satisfy the standard GMP algebra (\ref{eq: GMP}).

The projected Hamiltonian  is invariant under $U(1)$ and dipole symmetries generated by
\begin{align}
    Q = \rhob_0, \quad D_i = i\frac{\partial \rhob_{\bf k}}{\partial k_i} \Big|_{{\bf k}=0}.
\end{align}
These symmetries act on $\rhob_{\bf k }$  as
\begin{align}
    [Q, \rhob_{\bf k }]=0, \quad [D_i, \rhob_{\bf k }] =- \epsilon_{ij} k_j \rhob_{\bf k}.
\end{align}

The $U(1)$ symmetry is said to be broken if there exists a local operator $\phi({\bf x})$ called the order parameter such that
\begin{equation}
    \langle [Q,\phi({\bf x})]\rangle = C \ne  0.
\end{equation}
We assume that continuous translational symmetry is unbroken and thus the r.h.s. is $\bf x$-independent. 
After the Fourier transform we find
\begin{align}
    C \delta({\bf k}) = \frac{1}{(2\pi)^2} \int d^2{\bf y} \exp(-i {\bf k}\cdot{\bf y}) \langle [\rhob_{\bf k}, \phi_{-\bf k}] \rangle =\delta({\bf k}) \langle [\rhob_{\bf k}, \phi_{-\bf k} ]\rangle.
\end{align}
Since in position space $\langle [\rhob({\bf x}),\phi({\bf y})] \rangle$ decays at least exponentially for large separation $|\bf x - y|$, the Fourier transform\footnote{A cautious reader may notice that this expression is ill-defined in infinite volume. A more careful treatment is to work in real space and replace the charge $Q$ of infinite space with a charge computed in a large volume. The manipulations during the proof are almost the same as in \cite{KS} and we will omit the technical complication for sake of transparency. Alternatively, one can work in the momentum space of a finite volume system as was done in the original Hohenberg's paper \cite{Hohenberg} for example.  } $\langle [\rhob_{\bf k}, \phi_{-\bf k} ]\rangle$ is a smooth function near ${\bf k} = 0$ and we can divide this expression by $\delta(\bf k)$. We find 
\begin{align}
    \langle [\rhob_{\bf k}, \phi_{-\bf k} ]\rangle = C +o(|{\bf k}|).
\end{align}
Since the l.h.s is smooth there exists a region $|{\bf k}|<k_c$ such that $|\langle [\rhob_{\bf k}, \phi_{-\bf k} ]\rangle|>0$. We will restrict ourselves to this region in the following. 

We can apply to the l.h.s. the generalized uncertainty principle \cite{Pitaevskii1991}
\begin{equation*}
    |\langle [A_1,A_2]\rangle|^2 \le \langle A_1^\dagger A_1 +A_1A_1^\dagger\rangle \langle A_2^\dagger A_2 +A_2A_2^\dagger\rangle,
\end{equation*}
with $A_1=\phi_{-\bf k}$ and $A_2= \rhob_{{\bf k}}$. One  finds
\begin{equation}
    0<|\langle [\rhob_{\bf k}, \phi_{-\bf k} ]\rangle|^2 \le \langle  \rhob_{\bf k }^\dagger \rhob_{\bf k }+ \rhob_{\bf k } \rhob_{\bf k }^\dagger \rangle\langle  \phi_{-\bf k }^\dagger \phi_{-\bf k }+ \phi_{-\bf k }\phi_{-\bf k }^\dagger \rangle.
\end{equation}
Using Cauchy-Schwartz inequality we find
\begin{align}
\begin{split}
    \langle \rhob_{\bf k }^\dagger\rhob_{\bf k } \rangle = \int_0^\infty \frac {d\omega}{2\pi} \langle\rhob_{\omega{\bf k} }^\dagger\rhob_{\omega{\bf k} } \rangle \le \frac 1 2 \sqrt{\int_0^\infty \frac {d\omega}{\pi} \omega \langle\rhob_{\omega{\bf k} }^\dagger\rhob_{\omega{\bf k} } \rangle \int_0^\infty \frac {d\omega}{\pi}\frac 1 \omega \langle\rhob_{\omega{\bf k} }^\dagger\rhob_{\omega{\bf k}  }\rangle}\\=\frac 1 2 \sqrt{\langle [[\rhob^\dagger_{\bf k},\overline H],\rhob_{\bf k} ] \rangle}\sqrt{\chi_{\rhob\rhob}({\bf k})},
\end{split}
\end{align}
where we introduced time Fourier transform $\rhob_{\bf k}(t) = \int \frac {d\omega}{2\pi} \rhob_{\omega\bf k}e^{-i\omega t}$ and static compressibility $\chi_{\rhob\rhob}({\bf k})=\int_0^\infty \frac {d\omega}{\pi}\frac 1 \omega \langle\rhob_{\omega{\bf k} }^\dagger\rhob_{\omega{\bf k}  }\rangle$. Note that compressibility with respect to the projected densities is the same as compressibility with respect to unprojected ones. We will assume that  $\chi_{\rhob\rhob}({\bf k})$ is continuous function of $\bf k$ and approximate it with its value at ${\bf k}=0$. Moreover, we will assume that  the static compressibility is finite $\chi_{\rhob\rhob}(0) = \frac{\partial n}{\partial \mu} < \infty$, where $n$ is particle density and $\mu$ is chemical potential. For example, the FQHE  state is incompressible $\frac{\partial n}{\partial \mu} = 0$ and satisfies this condition.

Putting everything together, we find
\begin{align*}
    0<   \langle |\phi_{-\bf k}|^2\rangle \sqrt{\langle [[\rhob_{-\bf k },[\overline H,\rhob_{\bf k} ]] \rangle} \sqrt{\frac{\partial n}{\partial \mu}}.
\end{align*}
We assumed that all correlation functions are symmetric under reflection ${\bf k} \rightarrow - {\bf k}$ for simplicity of presentation. The general case can be done in the same fashion. 
\subsection*{Estimates} 

   We demand that
   \begin{align} \label{eq: phi for small k}
       \int d^2{\bf k} \,\langle |\phi_{\bf k}|^2\rangle  
   \end{align}
is finite. This condition is equivalent \cite{KS} to the requirement of cluster decomposition for $\phi$. Physically, it means that relative fluctuations of $\phi({\bf x})$ averaged over a region $U$ go to zero as the area of the region $|U|$ increases
\begin{align}
    \frac 1 {|U|^2} \int_{{\bf x},{\bf y}\in U} d^2 {\bf x} d^2 {\bf y} \, \Big(\langle\phi^\dagger({\bf x}) \phi({\bf y})\rangle - |\langle \phi({\bf x})\rangle|^2 \big)   \xrightarrow{|U|\rightarrow \infty} 0.
\end{align} 
This is a natural requirement if we assume that the order parameter is a well-defined quantity in a thermodynamical sense. Such assumption is implicit in the original proofs of HMW \cite{Hohenberg,MW}. 

In order for (\ref{eq: phi for small k}) to be finite, the $\langle |\phi_{\bf k}|^2\rangle $ cannot diverge too fast as momentum decreases: 

\begin{equation}
    \langle |\phi_{\bf k}|^2\rangle \sim |{\bf k}|^{\epsilon-2} \quad |{\bf k}|\rightarrow 0,
\end{equation}
with $\epsilon>0$. 

\bigskip

Next, we turn to computation of $\langle [[\rhob_{-\bf k },[\overline H,\rhob_{\bf k} ]] \rangle$. At this point, it is instructive to recall that before the LLL projection, one has the so-called f-sum rule
\begin{equation}
   \int_0^\infty \frac {d\omega}{\pi} \omega \langle\rho_{\omega{\bf k} }^\dagger\rho_{\omega{\bf k} } \rangle =\langle [[\rho_{-\bf k },[H,\rho_{\bf k} ]] \rangle = \frac {{\bf k}^2}{2m}
\end{equation}
After the projection, due to Kohn theorem \cite{Kohn} ${\bf k}^2$ part of the f-sum rule is saturated by cyclotron resonance which is projected out. After going to LLL, the remainder can be found by direct computation \cite{GMP}
\begin{align}\label{eq: sum rule}
   \langle [[\rhob_{-\bf k },[\overline H,\rhob_{\bf k} ]] \rangle \sim |{\bf k}|^4, \qquad |{\bf k}|\rightarrow 0,
\end{align}
but a more insightful way is to use symmetries. 
The low momentum expansion of density is
\begin{align}
    \rhob_{\bf k} \sim Q -i k_j D_j  + o(|{\bf k}|^2), \quad |{\bf k}|\rightarrow 0,
\end{align}
Since the Hamiltonian is invariant under charge and dipole symmetries the terms up third order will be zero. Physically, the dipole symmetry forces f-sum rule to start from quartic terms instead of the usual quadratic.   A more careful analysis shows that small ${\bf k}$ singularities in the second-order derivatives of the Fourier transform $V_{\bf k}$ of the interaction energy can contribute to the r.h.s. of (\ref{eq: sum rule}). The second-order derivatives of $V_{\bf k}$ are non-singular at the origin if the interaction energy $V({\bf x}-{\bf y})$ decays faster then the fourth power of separation $\dfrac{1}{|{\bf x}-{\bf y}|^4}$. Slower decays can invalidate our conclusions. 

Combining the estimates and assuming that $U(1)$ symmetry is broken we find
\begin{equation*}
        0<   \langle |\phi_{\bf k}|^2\rangle \sqrt{\langle [[\rhob_{-\bf k },[H,\rhob_{\bf k} ]] \rangle} \sqrt{\chi_{BB}} \sim |{\bf k}|^{\epsilon}  
\end{equation*}
which goes to $0$ as $|{\bf k}|\rightarrow 0$. This contradiction shows that $U(1)$ symmetry cannot be broken at zero temperature in the lowest Landau level.

\section{Inhomogeneous magnetic field}\label{sec: no homo} 
As was indicated in section \ref{sec: LLL}, the GMP algebra is still present even in the case of spatially varying magnetic field. However, the translational symmetry is lost and, as a consequence, the dipole symmetry. The latter is responsible for the absence of $k^2$ terms in the f-sum rule and the impossibility of symmetry breaking. In this section, we will take a simple toy model as an example and demonstrate this. In particular, we will explicitly compute the resulting GMP algebra and show that f-sum rule contains quadratic terms. 

We will consider a homogeneous magnetic field with a delta function insertion of one negative $2\pi$-flux at the origin
\begin{equation}
    B({\bf x}) = B_0 - 2 \pi \delta ({\bf x}).
\end{equation}

The ground state wave functions in the inhomogeneous magnetic field are of the form \cite{Solution}
\begin{equation}
    \psi({\bf x}) = f(z) \exp (-\phi(\bf x)),
\end{equation}
where we introduced complex coordinate notation $z = x_1+ i x_2$, $f(z)$ is a holomorphic function which grows not too fast at infinity,  $\phi({\bf x})$ satisfies
\begin{align}
    \Delta \phi ({\bf x}) =B({\bf x}),
\end{align}
and we used a symmetric gauge.

In our case, we find
\begin{align}
     \phi ({\bf x}) = \frac {|{\bf x}|^2}{4l_B^2} - \log |{\bf x}|,
\end{align}
and the normalized ground state wave functions are
\begin{align} \label{eq: basis in original gauge}
    \psi_n({\bf x}) = \frac {z^{n-1} |z|}{\sqrt{2\pi(2l_B^2)^{n}(n)! }}  \exp \Big(- \frac {|z|^2} {4 l_B^2}\Big), \qquad n\ge 1
\end{align}
where $l_b=\frac 1 {\sqrt{B_0}}$. All wave functions have a zero at ${z=0}$ and we can make a gauge transformation $\psi \rightarrow z \psi({\bf x})/|z|$. The resulting wave functions in the new gauge are
\begin{align}\label{eq: symmetric gauge wavefunctions}
    \psi_n({\bf x}) = \frac {z^{n} }{\sqrt{2\pi(2l_B^2)^{n}(n)! }}  \exp \Big(- \frac {|z|^2} {4 l_B^2}\Big), \qquad n\ge 1,
\end{align}
which are the same as the usual wave functions except that the wave function with $n=0$ is missing. It is reasonable to expect that since the Zeeman interaction leads to an infinite potential at the origin $|{\bf x}|=0$. 

The wave functions of the ground state are of the form  $\psi =  z f(z) \exp \Big(- \frac {|z|^2} {4 l_B^2}\Big)$ and one might be tempted to consider $f(z)$ which has a pole at the origin since the resulting wave function will be regular. However such functions should be excluded. Heuristically, the number of LLL states in a region is proportional to the total flux going through it. Since we have one unit of flux removed there should be one less state. More formally, the delta function flux at the origin should be physically thought of as a limit of insertion of a large negative magnetic field in a small area around ${\bf x}=0$. In this setting, the function $\phi({\bf x})$ does not have a logarithmic singularity which can cancel a pole in $f(z)$. One can show that in the zero size limit of the flux insertion area the wavefunctions reduce to (\ref{eq: basis in original gauge}). 

One can see that the Hilbert space of a system with a negative flux insertion can be thought of as a subspace of the homogeneous LLL with one state removed. We use that in the following and we will express the operators in the former Hilbert space as operators in the latter one which has zero matrix elements whenever one of the particles is in $n=0$ state. The computations are straightforward but cumbersome and we leave the details to the Appendix.

The Bergman kernel is 
\begin{align}
    \widetilde K({\bf x}, {\bf y})= \frac 1 {2 \pi l_B^2} \exp \Big(-\frac {{\bf x}^2}{4 l_B^2}-\frac {{\bf y}^2}{4 l_B^2}\Big)\left(\exp \Big( \frac {{\bf x}\cdot {\bf y}-i{\bf x}\wedge {\bf y}}{2l_B^2} \Big) -1  \right).
\end{align}
Here and in the following, we will use a tilde to distinguish the case with a flux inserted from a homogeneous magnetic field. The projected densities are
\begin{align}
    \widetilde \rho_{\bf k} = \rhob_{\bf k} - \exp\Big(-\frac{{\bf k}^2l_B^2}{2}\Big) \int \frac{l_B^2d^2{\bf q}}{2\pi} \left( 2 \exp\Big(\frac {l_B^2 {\bf k}\cdot{\bf q}} 2 \Big)\cos\Big(\frac {l_B^2 {\bf k}\wedge{\bf q}} 2 \Big)-1\right)\rhob_{\bf q},
\end{align}
where again $\rhob_{\bf k}, \widetilde  \rho_{\bf k}$ are projected densities for homogeneous and inhomogenous magnetic field  respectively. It satisfies 
\begin{align}\label{eq: rho tilde at 0}
   \int \frac{d^2{\bf k}}{(2\pi)^2}  \widetilde \rho_{\bf k} =0,
\end{align}
which is to be expected since this is $\widetilde \rho({\bf x})$ at ${\bf x}=0$.  But $\widetilde \rho_{\bf k}$ obey even stronger condition 
\begin{align}
   \int \frac{d^2{\bf k}}{(2\pi)^2} \exp \Big(\frac {l_B^2 {\bf q}\cdot{\bf k}} 2 \pm i\frac {l_B^2 {\bf q}\wedge{\bf k}} 2\Big)  \widetilde \rho_{\bf k} =0,
\end{align}
for any $\bf q$.

The GMP algebra is 
\begin{align}\label{eq: deformed GMP}
\begin{split}
    [\widetilde \rho_{\bf k}, \widetilde \rho_{\bf q}] &= 2i\, e^{{\frac {{\bf k}\cdot {\bf q}} 2} }\sin  \Big(\frac { {\bf k}\wedge {\bf q}} 2\Big)\, \widetilde\rho_{{\bf k}+{\bf q}}\\&+ 2i e^{-\frac{{\bf k}^2+{\bf q}^2+{\bf k}\cdot{\bf q}}{2}} \int \frac{d^2{\bf p}}{2\pi}\,e^{\frac { {\bf p}\cdot({\bf k}+{\bf q}) } 2 }\sin\Bigg(\frac {1} 2 \Big({\bf p}-\frac{{\bf k}+{\bf q}}2\Big)\wedge({\bf k}-{\bf q})\Bigg) \widetilde\rho_{\bf p},
\end{split}
\end{align}
where we set $l_B=1$ for simplicity. One can see that $ \widetilde \rho$ forms a closed algebra (which is a subalgebra of  homogeneous GMP algebra generated by $\rhob$). 

The dipole operator, defined as
\begin{equation}
    \widetilde D_i = i\frac{\partial \widetilde\rho_{\bf k}}{\partial k_i} \Big|_{{\bf k}=0},
\end{equation}
acts on densities as
\begin{align}
   [ \widetilde D_i,\widetilde \rho_{\bf q}] =- \epsilon_{ij} q_j\widetilde \rho_{\bf q}- \int \frac{d^2{\bf p}}{2\pi} e^{\frac { {\bf p}\cdot{\bf q}-{\bf q}^2 } 2 }\Bigg(p_i \sin \frac {{\bf q}\wedge{\bf p}} 2 -\epsilon_{ij} p_j \cos \frac {{\bf q}\wedge{\bf p}} 2\Bigg)\widetilde\rho_{\bf p}.
\end{align}
The latter term will lead to terms that are of the order  ${\bf k}^2$ in the sum rule (\ref{eq: sum rule}). For small $|{\bf k}|$, 
\begin{equation}
    [\widetilde \rho_{-\bf k}, [\overline H,\widetilde \rho_{\bf k}]] \sim  k_ik_j[D_i, [\overline H,D_j]]+\dots, \qquad |{\bf k}|\rightarrow 0
\end{equation}
and $[D_i, [\overline H,D_j]]$ is no longer equal zero on operator level. The resulting expression for the order ${\bf k}^2$ term is rather complicated, but we give a simple example for a translationally invariant state in the appendix. The ${\bf k}^2$ term makes $U(1)$ symmetry breaking possible in principle. 

An important difference from the homogeneous case is that there are no such vector-valued functions ${\bf h}({\bf x})$ that 
\begin{align*}
    h_i\star f -f\star h_i = \partial_i f \quad \text{for any $f$.}
\end{align*}
Indeed, if there were such a function one could translate the delta function at the origin $\delta({\bf x})$ into $\delta({\bf x}-{\bf a})$ using conjugation by a unitary which is exponentiation of the adjoint action of ${\bf a} \cdot {\bf h}$ (it is {\it not} equal to $\exp(i{\bf a} \cdot {\bf h})$). However, $\widetilde \rho(0)= \int d^2{\bf x}\, \delta({\bf x})\widetilde \rho({\bf x})$ is equal to 0 because of (\ref{eq: rho tilde at 0}) while    $\widetilde \rho({\bf a}) =\int d^2{\bf x}\, \delta({\bf x}-{\bf a})\widetilde \rho({\bf x})$ is not zero for general ${\bf a}$ and thus these operators cannot be related via a conjugation by a unitary. 

In a constant magnetic field, the corresponding function $\bf h$ is a linear function. One may wonder whether there is a remnant of dipole symmetry that descends from the translational invariance of the interactions before the projection. Or, in other words, whether the dipole symmetry gets deformed instead of disappearing.  An operator of a deformed symmetry should be of the form $\int d^2 {\bf x}\,\alpha_i({\bf x}) \widetilde \rho(\bf x)$ (where $\alpha_i({\bf x})$ is only linear on average) and should act as infinitesimal translation on $\widetilde \rho(\bf x)$. The argument of the previous paragraph shows that this is impossible and dipole symmetry is lost.

Let us conclude this section with a couple of straightforward generalizations. First, one can add $N$ negative $2\pi$ fluxes at the origin instead of just one flux. This will remove the $n=0,\dots N-1$ wave functions (\ref{eq: symmetric gauge wavefunctions}). It is an interesting question what happens to GMP algebra in the limit when $N$ is large but finite. In particular, can the algebra be approximated as a sum of bulk and boundary components?  Secondly, one can remove $2\pi$ fluxes forming some lattice $\Lambda$ by subtracting $\sum_{{\bf R },{\bf R' }\in \Lambda} \psi_{\bf R}({\bf x})K_{\Lambda}^{-1}({\bf R },{\bf R' })\psi^\dagger_{\bf R'}({\bf y})$ from the homogeneous kernel. Here $\psi_{\bf R }({\bf x}) = K({\bf x},{\bf R}) $ is a coherent state which corresponds to a wave function localized around $\bf R$ and $K_{\Lambda}^{-1}({\bf R },{\bf R' })$ is inverse to $K({\bf R },{\bf R' })$ understood as an infinite matrix labeled by discrete indices ${\bf R },{\bf R' }\in \Lambda$, i.e. $\sum_{{\bf R''}\in \Lambda}K({\bf R },{\bf R'' })K_{\Lambda}^{-1}({\bf R'' },{\bf R' })=\delta_{{\bf R },{\bf R '}}$. The inverse matrix $K_{\Lambda}^{-1}({\bf R },{\bf R' })$ is hard to compute, but one can do it approximately by assuming that periods of the lattice $\Lambda$ are much greater than  the magnetic length. One can compute the resulting approximate GMP algebra.

\section{Discussion}

Let us  discuss the limitation of the theorem. First, at zero temperature, the condition of finite compressibility $\frac {\partial n}{\partial \mu} < \infty$ is important with free boson theory being a counter-example. Another important counter-example is the quantum Hall ferromagnet, where the relevant susceptibility diverges since it will just reorient the vacuum magnetization. 

Secondly, the argument does not forbid condensation of particles as long as they are neutral under $U(1)$. In particular, the Moore-Read state \cite{MooreRead} viewed as a superconducting state of composite fermions is not forbidden. The composite fermion particle number is coupled to an emergent gauge field which can undergo the Higgs transition and avoid the HMW argument. 

A related issue is long-range interactions. One has to assume that the interactions decay fast enough since otherwise there could be singularities in the Fourier transform of the two-body interaction $V_{\bf k}$ which will affect the sum rule. 

Lastly, the higher Landau levels can lead to the breaking of $U(1)$ symmetry but the relevant effect will be suppressed by the gap. One should note that heating the system to a small non-zero temperature typically will lead to enhanced fluctuations of the order parameter and further increase the tendency to symmetry restoration.

Even though spontaneous symmetry breaking is absent and the existence of massless excitations is not enforced by the Goldstone theorem, one can still write down the would-be effective theory pretending that the symmetry is broken. It can still be useful in understanding for example the Berezinskii–Kosterlitz–Thouless transitions in this system. For the $U(1)$ symmetry breaking the relevant theory should be a non-commutative field theory similar to the theory Tkachenko mode \cite{tkachenko} arising in the rotating superfluid  vortex lattice. It is an interesting problem what happens in the effective theory after the introduction of long-range Coulomb interaction. The long-range interaction could gap the would-be Goldstone modes and prevent their fluctuations from restoring the $U(1)$ symmetry. The resulting theory may have relevance to the superconductivity in magic-angle twisted bilayer graphene.

\section*{Appendix}
In this appendix, we will set $l_B = 1$ and use a tilde to distinguish a case with a flux insertion from a homogeneous one. Everything without a tilde is assumed to be computed for a homogeneous field. We will use the  gauge where (\ref{eq: symmetric gauge wavefunctions}) holds.
\subsection{Homogeneous field}
Let us first review the computation in the constant magnetic field.  The Bergman kernel is
\begin{align}
   K({\bf x}, {\bf y})= \sum_{n\ge0} \psi_n({\bf x})\psi_n^\dagger({\bf y})=\frac 1 {2 \pi } \exp \Big(-\frac {{\bf x}^2}{4 }-\frac {{\bf y}^2}{4 }\Big)\exp \Big(  \frac {{\bf x}\cdot {\bf y}-i{\bf x}\wedge {\bf y}}{2} \Big)
\end{align}
where $\psi_n$ are given by (\ref{eq: symmetric gauge wavefunctions}). 

It is convenient to write single-particle operators in the LLL in terms of their kernels 
\begin{equation}
    A({\bf x},{\bf y})= \sum_{n,m\ge 0} \psi_n({\bf x})\langle n | \hat A|m\rangle \psi_m^\dagger({\bf y}).
\end{equation}
It is straightforward to show that
\begin{align}
(\hat A \psi) ({\bf x}) &= \int d^2{\bf y} \, A({\bf x},{\bf y}) \psi({\bf y}),\\
\int d^2{\bf z}  \, K({\bf x},{\bf z}) A({\bf z},{\bf y}) &= \int d^2{\bf z}  \, A({\bf x},{\bf z}) K({\bf z},{\bf y}) =  A({\bf x},{\bf y}),\\ 
(\hat A \hat B) ({\bf x},{\bf y}) &= \int d^2{\bf z}  \, A({\bf x},{\bf z}) B({\bf z},{\bf y}),\\
{\rm Tr} \,\hat A &= \int d^2{\bf x}  \, A({\bf x},{\bf x}),\\
 T_f ({\bf x},{\bf y}) &= \int d^2{\bf z}  \, K({\bf x},{\bf z}) f({\bf z}) K({\bf z},{\bf y}),
\end{align}
where $\hat T_f$ is a projection of multiplication by function $f({\bf x})$ to the LLL.

In the homogeneous magnetic field, the magnetic translation operators \cite{Zak} act on wavefunctions as
\begin{align}
    \hat {\mathcal T}_{\bf a}\psi({\bf x}) = \exp \Big( \frac{i}{2} {\bf x }\wedge{\bf a}\Big)\psi({\bf x}+ {\bf a}).
\end{align}
They commute with the Hamiltonian for all ${\bf a}$ and thus with the projection
\begin{align}
   \hat {\mathcal T}_{\bf a}\hat K = \hat K  \hat {\mathcal T}_{\bf a},
\end{align}
or in terms of kernel
\begin{align}
   \exp \Big( \frac{i}{2} {\bf x }\wedge{\bf a}\Big)K({\bf x+a},{\bf y}) =\exp \Big( \frac{i}{2} {\bf y}\wedge{\bf a}\Big)K({\bf x},{\bf y-a}).
\end{align}

The magnetic translations satisfy the magnetic translation algebra before and after the projection
\begin{equation}\label{eq: mag trans algebra}
   \hat {\mathcal T}_{\bf a} \hat {\mathcal T}_{\bf b} = \exp\left(  i { {\bf a }\wedge {\bf b}}\right)   \hat {\mathcal T}_{\bf b} \hat {\mathcal T}_{\bf a}.
\end{equation}
They act on Toeplitz operators $\hat T_f$ as
\begin{align} \label{eq: mag action on toeplitz}
   \hat {\mathcal T}_{\bf a} \hat T_{f}\hat {\mathcal T}_{-\bf a} = \hat T_{f^{\hat {\mathcal T}_{\bf a}}}, \qquad \text{where $f^{\hat {\mathcal T}_{\bf a}} ({\bf x})= f({\bf x}+{\bf a})$}.
\end{align}

The projection of the magnetic translation operator is expressible as Toeplitz operator $\hat T_{f_{\bf a}}$ for some function $f_{\bf a}({\bf x})$. From the magnetic translation algebra (\ref{eq: mag trans algebra}) we find
\begin{equation}
  \hat {\mathcal T}_{\bf b}   \hat {\mathcal T}_{\bf a} \hat {\mathcal T}_{-\bf b} = \exp\left(  -i{ {\bf a }\wedge {\bf b}}\right)  \hat {\mathcal T}_{\bf a}.
\end{equation}
By comparing it with (\ref{eq: mag action on toeplitz}), we find that up to $\bf x$-independent prefactor $f_{\bf a}({\bf x})\sim\exp \left(i{\bf x}\wedge{\bf a}\right)$. The prefactor can be found by projecting $\exp \left(i{\bf x}\wedge{\bf a}\right)$  and comparing it with the magnetic translation. One finds for the exponential functions $e_{\bf k} ({\bf x })= \exp(-i {\bf k}\cdot {\bf x})$
\begin{align}\label{eq: proj exp}
    T_{e_{\bf k}} ({\bf x },{\bf y}) =  \exp \Big(-\frac {{\bf k}^2 } 2+\frac {{\bf k}\wedge ({\bf x}-{\bf y}) } 2 -\frac {i{\bf k}\cdot({\bf x}+{\bf y}) } 2\Big)K({\bf x},{\bf y}).
\end{align}

And we find
\begin{equation}\label{eq: trans in terms of functions}
    f_{\bf a}({\bf x}) = \exp \left(\frac {{\bf a}^2} {4} +i{\bf x}\wedge{\bf a}\right).
\end{equation}
Thus magnetic translation operators $\hat {\mathcal T}_{\bf a}$ are
\begin{align}\label{eq: mag trans in terms of Toeplitz}
    \hat {\mathcal T}_{\bf a} = \hat T_{f_{\bf a}}.
\end{align}
We can find a nice basis of unitary operators 
\begin{align}
    \hat \tau_{\bf k} = \hat {\mathcal T}_{-\wedge {\bf k}} =  \exp \Big(\frac {{\bf k}^2 } 4 \Big)\hat T_{e_{-\bf k}}.
\end{align}
Labeling the basis in terms of $\bf k$ instead of ${\bf a}=-{\wedge {\bf k}}$ is conventional and chosen here in order to reduce to the standard Fourier basis $\exp(i {\bf k}\cdot {\bf x})$ in $l_B\rightarrow 0$ limit. The kernels of these operators are 
\begin{align}\label{eq: proj exp}
\tau_{\bf k} ({\bf x },{\bf y}) =  \exp \Big(-\frac {{\bf k}^2 } 4-\frac {{\bf k}\wedge ({\bf x}-{\bf y}) } 2 +\frac {i{\bf k}\cdot({\bf x}+{\bf y}) } 2\Big)K({\bf x},{\bf y}),
\end{align}
and they satisfy the identity

\begin{align}
  \hat   \tau_{\bf k}   \hat   \tau_{\bf q} =\exp\Big(i\frac{{\bf k}\wedge{\bf q}}2\Big)\tau_{{\bf k}+{\bf q}}.
\end{align}
They form a basis for the space of operators
\begin{align} \label{eq: tau orthognality}
\begin{split}
\hat \tau^\dagger_{\bf k}&=\hat \tau_{-\bf k},\\
{\rm Tr}\,\Big (\hat \tau_{\bf k}\hat \tau_{\bf p}\Big)  &= 2\pi\delta({\bf k}+{\bf p}),\\
    \int \frac{d^2 {\bf k}}{2\pi} \,\langle n | \hat \tau_{\bf k} |m\rangle \langle l | \hat \tau_{-\bf k} |s \rangle  &=\delta_{ns}\delta_{ml},
\end{split}
\end{align}
where $|n\rangle$ is ket notation for the orthonormal basis $\psi_n$. Any operator can be expanded in this basis 
\begin{align}
    \hat A  = \int \frac{d^2 {\bf k}}{2\pi} \,A_{{\bf k}} \hat \tau_{\bf k},
\end{align}
where the Fourier components $A_{{\bf k}}$ are defined as 
\begin{align}
    A_{\bf k} = {\rm Tr } (\hat A \hat \tau_{-{\bf k}})  = \int d^2 {\bf x}d^2{\bf y} \,A({\bf x},{\bf y}) \tau_{-\bf k}({\bf y},{\bf x}).
\end{align}
Note a useful identity 
\begin{align}
    \hat A = \hat T_{f_{\hat A}},\qquad \text{where $f_{\hat A}({\bf x})= \int \frac {d^2{\bf k}}{2\pi} A_{\bf k} \exp\Big({\frac{{\bf k}^2}4+i {\bf k}\cdot {\bf x}}\Big)$}.
\end{align}
This is one of the main formulas we were aiming for. It allows the expression of an arbitrary operator in the lowest Landau level in terms of projected multiplication by a function $\hat T_f$. 

For example, we find 
\begin{align}
    {\rm Tr}\big([\hat T_{e_{\bf k}},\hat T_{e_{\bf q}}] \hat \tau_{-{\bf p}}\big) = 4\pi i\, \exp \Big(-{\dfrac {{\bf k}^2+{\bf q}^2} 4} \Big)\sin  \Big(\frac { {\bf k}\wedge {\bf q}} 2 \Big) \delta({\bf p }+{\bf k }+ {\bf q}),
\end{align}
which leads to
\begin{align}\label{eq: commutator of toeplitz}
    [\hat T_{e_{\bf k}},\hat T_{e_{\bf q}}] = 2i\, \exp \Big({\dfrac {{\bf k}\cdot {\bf q}} 2} \Big)\sin  \Big(\frac { {\bf k}\wedge {\bf q}} 2 \Big)\, \hat T_{e_{{\bf k}+{\bf q}}}.
\end{align}
Using the identity
\begin{align}
    \rhob(f) = \int d^2{\bf x} d^2{\bf y}\,\psib^\dagger({\bf x}) T_f({\bf x},{\bf y}) \psib^\dagger({\bf y}) 
\end{align}
together with (\ref{eq: star product def}) and (\ref{eq: commutator of toeplitz}), we find the standard GMP algebra (\ref{eq: GMP}).

In this subsection, we took a long path that extensively utilizes the continuous magnetic translation symmetry in order to find Toeplitz operators corresponding to any projected operator and compute their algebra. This way once again emphasizes the importance of symmetry in the computability of the standard GMP algebra. One can use a similar method for any symmetric space.  In the case of an inhomogeneous system, the symmetry disappears along with the nice basis $\hat \tau_{\bf k}$. One can still project the plane waves and find $\hat T_{e_{\bf k}}$, but they won't be orthogonal to each other in the sense of eqs. (\ref{eq: tau orthognality}).

This machinery allows a straightforward computation of GMP algebra in the case of a homogeneous field with a negative flux insertion.

\subsection{Homogeneous field with a negative flux}
The kernel can be found as
\begin{align}
\begin{split}
   \widetilde K({\bf x}, {\bf y})= \sum_{n\ge1} \psi_n({\bf x})\psi_n^\dagger({\bf y})&= K({\bf x}, {\bf y}) -\psi_0({\bf x})\psi_0^\dagger({\bf y})\\&=\frac 1 {2 \pi } \exp \Big(-\frac {{\bf x}^2}{4 }-\frac {{\bf y}^2}{4 }\Big)\left(\exp \Big(  \frac {{\bf x}\cdot {\bf y}-i{\bf x}\wedge {\bf y}}{2} \Big) -1  \right).
\end{split}
\end{align}
As obvious from the second equality, we can think about projection to the system with a negative flux as first projecting to the homogeneous LLL and then projecting out a single state $n=0$. We will use that in the following and express the operators acting in the $\widetilde {\mathcal H}_{LLL}$ as operators acting in the $\mathcal H_{LLL}$ with zero matrix elements with $n=0$ state. 

We can express the kernel in terms of the basis using
\begin{align}
     {\rm Tr} \big(\widehat {\widetilde K} \hat \tau_{-{\bf q}}\big) = 2\pi \delta({\bf q}) - \exp\left({-\frac {\bf q^2}4}\right).
\end{align}
By expressing $\widetilde K$ in terms of $\tau_{\bf q}$ the computation of the GMP algebra becomes a straightforward exercise.

The projection of exponential functions gives
\begin{align*}
    {\rm Tr} \big(\widehat {\widetilde K} \hat T_{e_{\bf k}}\widehat {\widetilde K} \hat \tau_{-{\bf q}}\big)=2\pi \exp\Big(-\frac{k^2}4\Big)\delta({\bf k}+{\bf q}) - \exp\Big(-\frac{{\bf k}^2}{2}-\frac{{\bf q}^2}{4}\Big) \left( 2 \exp\Big(-\frac { {\bf k}\cdot{\bf q}} 2 \Big)\cos\Big(\frac {{\bf k}\wedge{\bf q}} 2 \Big)-1\right),
\end{align*}
or in terms of density operators
\begin{align}\label{eq: rho tilde}
    \widetilde \rho_{\bf k} = \rhob_{\bf k} - \exp\Big(-\frac{{\bf k}^2}{2}\Big) \int \frac{d^2{\bf q}}{2\pi} \left( 2 \exp\Big(\frac { {\bf k}\cdot{\bf q}} 2 \Big)\cos\Big(\frac { {\bf k}\wedge{\bf q}} 2 \Big)-1\right)\rhob_{\bf q}.
\end{align}

Note, if we act on both sides of (\ref{eq: rho tilde}) with projections $\widetilde P_{LLL}$ it must become a trivial identity. In order for this to be true $\widetilde\rho_{\bf q}$ must satisfy
\begin{align}
    \int \frac{d^2{\bf q}}{2\pi}  \exp\Big(\frac { {\bf k}\cdot{\bf q}} 2 \Big)\cos\Big(\frac {l_B^2 {\bf k}\wedge{\bf q}} 2 \Big) \widetilde\rho_{\bf q} = 0,
\end{align}
which can be checked directly. Actually, a stronger condition  holds. It originates from the fact that similar self-consistency conditions must hold when one expresses $\widetilde \rho_{\bf k}$ in terms of operators $\widetilde \rho_{\bf k}^{L}$ and $\widetilde \rho_{\bf k}^{R}$ which are $\rhob_{\bf k}$ projected by  $\widetilde P_{LLL}$ only on the left or on the right respectively. The resulting conditions are
\begin{align}\label{eq: n=0 conditon}
   \int \frac{d^2{\bf q}}{2\pi} \exp \Big(\frac { {\bf q}\cdot{\bf k}} 2 \pm i\frac { {\bf q}\wedge{\bf k}} 2\Big)  \widetilde \rho_{\bf q} =0, \qquad \text{for any $\bf k$}
\end{align}
whose again can be checked directly. As clear from the above discussions, these conditions enforce matrix elements of $\widetilde \rho$ to be zero whenever one of the particles is in $n=0$ state.

Lastly, we want to compute the GMP algebra. There are two ways to do it. One is to compute ${\rm Tr} \big( [\widehat {\widetilde T}_{e_{\bf k}},\widehat {\widetilde T}_{e_{\bf q}}] \hat \tau_{-{\bf p}}\big)$, where $\widehat {\widetilde T}_{e_{\bf k}}=\widehat {\widetilde K} \hat T_{e_{\bf k}}\widehat {\widetilde K}$. The other way is to use (\ref{eq: rho tilde}) together with standard GMP algebra relations (\ref{eq: GMP}) and constraints (\ref{eq: n=0 conditon}). Either computation is lengthy but straightforward. The result is (\ref{eq: deformed GMP}).

Consider a translationally and rotationally invariant state (even though the Hamiltonian is not invariant)
\begin{align}
\langle \widetilde \rho_{\bf p}\widetilde \rho_{\bf q}\rangle = (2\pi)^2s_{\bf q}\delta({\bf p}+ {\bf q}),
\end{align}
where we introduced the structure factor $s_{\bf q}$ which depends only on $|\bf q|$. In this state one finds

\begin{equation}
   \sum_i\langle [D_i, [\overline H,D_i]]\rangle= \int\frac{d^2 {\bf k}}{\pi}  4\pi{\bf k}^2 e^{\frac{-{\bf k}^2}2}V_{\bf k} s_{\bf k}+\int\frac{d^2 {\bf k}d^2 {\bf p}}{(2\pi)^4}{\bf k}^2e^{{\bf p}\cdot {\bf k}{-{\bf p}^2}} V_{\bf k} s_{\bf p},
\end{equation}
which is not zero for general interaction. 

\bibliographystyle{apsrev4-1}
\bibliography{bib}
\end{document}